\documentclass[a4paper,11pt]{article}
\usepackage[bbgreekl]{mathbbol}
\usepackage{subfig} 
\usepackage{caption}
\usepackage[small]{titlesec}
\usepackage[utf8]{inputenc}
\usepackage[T1]{fontenc}  
\usepackage[backend=biber,style=numeric-comp,natbib=false,giveninits=true,doi=true,isbn=false,url=false,date=year,sorting=nyvt]{biblatex}
\DeclareNameAlias{default}{family-given}

\DeclareFieldFormat[article]{title}{#1}
\renewbibmacro{in:}{}
\DeclareFieldFormat[article]{pages}{#1}
\DeclareFieldFormat{journal}{#1}
\renewcommand\bf\bfseries

\renewbibmacro*{volume+number+eid}{%
	\printfield{volume}%
	\setunit*{\addnbspace}
	\printfield{number}%
	\setunit{\addcomma\space}%
	\printfield{eid}}
\DeclareFieldFormat[article]{number}{\mkbibparens{#1}}
\DeclareFieldFormat[article]{volume}{\textbf{#1}}
\DeclareFieldFormat{year}{\mkbibparens{#1}}
\DeclareBibliographyDriver{article}{%
	\printnames{author}:%
	\newunit\newblock
	\printfield{title}%
	\newunit\newblock
	\printfield{journaltitle}
	\newunit
	\iffieldundef{number}{\printfield{volume}}{\printfield{volume}\addspace\printfield{number}}
	\addcomma\addspace\printfield{pages}\addspace
	\printfield{year}
}
\AtEveryBibitem{\clearfield{month}}
\AtEveryBibitem{\clearfield{day}}
\usepackage[english]{babel}
\usepackage[T1]{fontenc}
\usepackage{csquotes}
\usepackage{bbm}
\usepackage{amsmath,amsfonts,amsthm,amsbsy,amssymb,dsfont,stmaryrd}
\usepackage[dvipsnames]{xcolor}
\usepackage{braket}

\usepackage{mathtools}

\numberwithin{equation}{section}

\newcommand\myshade{85}
\colorlet{mylinkcolor}{violet}
\colorlet{mycitecolor}{YellowOrange}
\colorlet{myurlcolor}{Aquamarine}

\usepackage[left=2.5cm,right=2.5cm,top=2.5cm,bottom=2.5cm]{geometry}
\usepackage[unicode=true,pdfusetitle,
bookmarks=true,bookmarksnumbered=false,bookmarksopen=false,
breaklinks=false,pdfborder={0 0 1},backref=false,
linkcolor  = mylinkcolor!\myshade!black,
citecolor  = mycitecolor!\myshade!black,
urlcolor   = myurlcolor!\myshade!black,
colorlinks = true,
]
{hyperref}
\usepackage[nameinlink]{cleveref}

\usepackage{subfig}
\usepackage{tikz}
\usetikzlibrary{arrows}
\usetikzlibrary{intersections}

\usepackage{graphicx}
\usepackage{caption}

\definecolor{ct_black}{HTML}{000000}
\definecolor{ct_orange}{HTML}{ED872D}
\definecolor{ct_purple}{HTML}{7A68A6}
\definecolor{ct_blue}{HTML}{348ABD}
\definecolor{ct_turquoise}{HTML}{188487}
\definecolor{ct_red}{HTML}{E32636}
\definecolor{ct_pink}{HTML}{CF4457}
\definecolor{ct_green}{HTML}{467821}

\definecolor{ct2_green}{HTML}{9FF781}
\definecolor{ct2_green_dark}{HTML}{088A08}

\theoremstyle{plain}
\newtheorem{thm}{\protect\theoremname}[section]
\theoremstyle{plain}

\theoremstyle{plain}

\theoremstyle{plain}

\theoremstyle{remark}

\newtheorem{assumption}[thm]{\protect\assumptionname}

\theoremstyle{remark}
\newtheorem{rem}[thm]{\protect\remarkname}
\theoremstyle{definition}

\theoremstyle{plain}

\providecommand{\assumptionname}{Assumption}
\providecommand{\claimname}{Claim}
\providecommand{\corollaryname}{Corollary}
\providecommand{\definitionname}{Definition}
\providecommand{\lemmaname}{Lemma}
\providecommand{\propositionname}{Proposition}
\providecommand{\remarkname}{Remark}
\providecommand{\theoremname}{Theorem}
\providecommand{\examplename}{Example}

\crefname{section}{Section}{Sections}
\crefname{appendix}{Appendix}{Appendices}
\crefname{figure}{Figure}{Figures}
\crefname{assumption}{Assumption}{Assumptions}
\crefname{thm}{Theorem}{Theorems}
\crefname{lem}{Lemma}{Lemmas}
\crefformat{equation}{(#2#1#3)}

\crefrangelabelformat{equation}{(#3#1#4--#5#2#6)}

\crefmultiformat{equation}{(#2#1#3}{, #2#1#3)}{#2#1#3}{#2#1#3}
\Crefmultiformat{equation}{(#2#1#3}{, #2#1#3)}{#2#1#3}{#2#1#3}

\newtheorem*{lem*}{\protect\lemmaname}

\newcommand{\ee}{\operatorname{e}}
\newcommand{\ii}{\operatorname{i}}

\newcommand{\ZZ}{\mathbb{Z}}

\newcommand{\HSSH}{H_{_{\rm SSH}}}
\newcommand{\tin}{t_{_{\rm in}}}
\newcommand{\tout}{t_{_{\rm out}}}
\newcommand{\din}{d_{_{\rm in}}}
\newcommand{\dout}{d_{_{\rm out}}}
\newcommand{\wA}{w_A}
\newcommand{\wB}{w_B}

\newcommand{\GG}{\mathbb{G}}
\newcommand{\RR}{\mathbb{R}}
\newcommand{\CC}{\mathbb{C}}

\newcommand{\norm}[1]{\left\|#1\right\|}
\newcommand{\ip}[2]{\langle #1, #2 \rangle}

\newcommand{\ve}{\varepsilon}
\newcommand{\vf}{\varphi}
\newcommand{\Id}{\mathds{1}}
\newcommand{\HH}{\mathcal{H}}

\usepackage{environ}

\NewEnviron{malign}{%
	\begin{align}\begin{split}
			\BODY
	\end{split}\end{align}
}

\newcommand{\eq}[1]{\begin{align*}#1\end{align*}}
\newcommand{\eql}[1]{\begin{align}#1\end{align}}

\title{Is the continuum SSH model topological?}
\author{Jacob Shapiro\\
	{\footnotesize Department of Physics, Princeton University}\\
	 Michael I. Weinstein\\
		\footnotesize{Department of Applied Physics and Applied Mathematics, and Department of Mathematics, Columbia University}
}

\begin{document}
	
\maketitle

\begin{abstract}

The discrete Hamiltonian of Su, Schrieffer and Heeger (SSH) \cite{SSH_1979} is a well-known one-dimensional translation-invariant model in condensed matter physics. The model consists of two atoms per unit cell and describes in-cell and out-of-cell electron-hopping between two sub-lattices. It is among the simplest models exhibiting a non-trivial topological phase; to the SSH Hamiltonian one can associate a winding number, the Zak phase, which depends on the ratio of hopping coefficients and takes on the values $0$ and $1$ labeling the two distinct phases. We display two homotopically equivalent continuum Hamiltonians whose tight binding limits are SSH models with different topological indices. The topological character of the SSH model is therefore an emergent rather than fundamental property, associated with emergent chiral or sublattice symmetry in the tight-binding limit.

In order to establish that the tight-binding limit of these continuum Hamiltonians is the SSH model, we extend our recent results on the tight-binding approximation \cite{Shapiro_Weinstein_2020} to lattices which depend on the tight-binding asymptotic parameter $\lambda$.
\end{abstract}
 
\section{Introduction}
A central goal of condensed matter physics is an understanding of which materials are conducting and which are insulating. From the study of the integer quantum Hall effect (IQHE) arose an understanding that there are classes of materials, which in select energy ranges, have robust bulk and boundary conduction properties which can be understood via the algebraic topology of the space of quantum mechanical Hamiltonians. In the field of {\it topological insulators} different materials (or the same material at different phases) are associated with the distinct connected components in a space of Hamiltonians with an energy gap. Associated with each of these components is an integer-valued invariant, which labels the "phases" of the material. Materials (Hamiltonians) in two distinct components cannot be deformed into one another without closing the energy gap.

The development of the subject is largely based on tight binding models, discrete models obtained
via orbital or Wannier function approximations, which are a powerful approximate tool for describing quantum systems in restricted energy ranges. The goal of this paper is to illustrate, in the context of a particular tight-binding model, the well-known SSH model, that its topological character is a property that is merely emergent in the tight-binding limit. In particular, we display two homotopically equivalent continuum Hamiltonians, whose tight binding limits are in distinct topological classes. Hence, topology in the SSH model is an emergent, rather than fundamental, property. It is associated with the emergence of chiral or sublattice symmetry in the tight-binding limit.

\bigskip

\subsection{The SSH model and the goal of this paper}\label{ssh}

The Su-Schrieffer–Heeger (SSH) model \cite{SSH_1979} was initially proposed to model polyacetelyne, a molecular chain of alternating carbon and hydrogen atoms with alternating single and double bonds.
An idealization is a one-dimensional chain; the line is partitioned into cells, each containing two sites; in the $n^{th}$ cell are $A_n$ ("carbon") followed by $B_n$ ("hydrogen").
%
%
To these sites, we assign complex amplitudes $\psi^A_n$ and $\psi^B_n$. The wave function for cell $n$ is then $\psi_n=(\psi^A_n,\psi^B_n)^\top\in\CC^2$. Hence, the SSH wave function is regarded as $\psi=\{\psi_n\}_{n\in\ZZ}\in l^2(\ZZ;\CC^2)$. The SSH Hamiltonian $\HSSH:l^2(\ZZ;\CC^2)\to l^2(\ZZ;\CC^2) $ is defined by:
\begin{equation} \left(\HSSH\psi\right)_n =  \begin{pmatrix} (\HSSH\psi)^A_n \\ (\HSSH\psi)^B_n \end{pmatrix}
 = 
\begin{pmatrix}
\tin\ \psi^B_n + \tout\ \psi^B_{n-1} \\
\overline{\tin}\ \psi^A_n + \overline{\tout}\ \psi^A_{n+1} \end{pmatrix},\quad n\in\ZZ; 
\label{HSSH}\end{equation}
an electron hops between nearest neighbor sites. The in-cell hopping coefficient is denoted $\tin$
 and the out-of-cell hopping coefficient is denoted $\tout$. The band structure of $\HSSH$ is easily obtained. We seek non-trivial plane wave solutions, $\Psi=\{\psi_n\}$, bulk spectral problem: $\HSSH\Psi=E\Psi$. These take the form $
 \Psi_n = e^{ikn}\xi,\quad \xi=(\xi^A,\xi^B)^\top\in\CC^2$ and $n\in\ZZ$. Thus, $(\HSSH(k)-E)\xi=0$, 
 with $\xi\ne0$, where
\begin{equation}
 \HSSH(k) = \begin{bmatrix} 0 & s(k) \\ \overline{s(k)} & 0 \end{bmatrix},\quad 
s(k) =  \tin + \tout\exp(-\ii k).
\label{HSSHk}\end{equation}
Thus, $\HSSH$ has two dispersion curves $k\mapsto E_\pm(k)=\pm|s(k)|,\ k\in[0,2\pi].$
 A spectral gap exists if and only if $\tin\ne \tout$, and for $\tin\ne \tout$ the spectral
  gap has width $2|\tin-\tout|$. The two "phases" of $\HSSH$ are associated with the case
   $|\tin/\tout|<1$ and $|\tin/\tout|>1$ for which the winding number about the origin of the map:
  \[ [0,2\pi]\ni k \mapsto s(k)\in \CC\setminus\Set{0} \]
  is zero or one, respectively. Physically, this number is associated with the signed (by chirality) number of edge modes in a half-infinite sample, or a chiral polarization in the bulk \cite{Prodan_Song_2014,Graf_Shapiro_2018_1D_Chiral_BEC}.
 
 This non-trivial topological character is related to the off-diagonal structure \cref{HSSHk}, often referred to as "chiral" or "sub-lattice" symmetry. This is consistent with Kitaev's classification of Hamiltonians according to symmetry class and dimensionality \cite[Table 1]{Hasan_Kane_2010}, where one-dimensional models with no symmetry constraints are topologically trivial (class A) and chiral models are indexed by a $\ZZ$-valued invariant (class AIII).
 More intuitively, for there to be non-trivial topology in 1D, the complete one-dimensional Anderson localization \cite{Carmona1987} has to fail somehow, and in the SSH model, that may happen at zero energy thanks to chiral symmetry \cite{Graf_Shapiro_2017}. See e.g. \cite{Germinet_Klein_Schenker_2007} for an explanation of this mechanism in the context of the IQHE.
 
It is natural to ask whether the topological properties of 1D discrete models are present in the continuum models of which these discrete models are approximations. For the case of continuum two-dimensional crystals in a strong constant magnetic field, the setting of the integer quantum Hall effect (IQHE), we recently proved the equality of topological invariants for continuum Hamiltonians -- in the strong binding regime -- with those of the discrete tight binding limit \cite{Shapiro_Weinstein_2020}.

\medskip
\noindent{\it Question: Do discrete chiral models arise as the tight-binding limit of a \emph{topologically non-trivial} continuum model, just as for the 2D IQHE?}
\medskip

To be sure, in the continuum setting, chiral topological models do exist according to the various classification tables in the continuum \cite{Kubota2017,Bourne2017}, and one may even write topological invariants for these models. However, they are phrased on the already-graded Hilbert space $L^2(\RR)\otimes\CC^2$ and suggest more the Dirac equation rather than the Schroedinger equation. Hence it is not clear to us that there is actually a topologically non-trivial continuum one-dimensional Schroedinger particle. Another point of tension is due to the fact that chiral symmetry implies the energy spectrum is symmetric about zero, whereas the spectrum of a reasonable continuum Schroedinger operator is half-infinite.

In this paper we display two homotopically equivalent continuum Hamiltonians whose tight binding limits are SSH models with different topological indices. Therefore, while no topological invariant
 distinguishes between our continuum Hamiltonians, the two discrete limiting Hamiltonians are topologically distinct. The topological character of the SSH model is therefore an emergent property, arising only in the tight-binding limit.
 
A point of independent interest in the present work is that in order to obtain convergence to a limiting tight-binding model for\ our continuum models, in the limit where the depth of the wells, $\lambda$, tends to infinity,  we need an extension of the technique developed in \cite{Shapiro_Weinstein_2020} to the situation where the underlying lattice itself depends on the asymptotic parameter $\lambda$. This is achieved in the next section.

\medskip

\begin{rem}
The arguments of \cite{Shapiro_Weinstein_2020} establish norm resolvent convergence of the continuum Hamiltonian to the tight-binding (discrete) limit, a notion of convergence enabling the control of topological indices for IQHE, which are expressed in terms
 of spectral (Fermi) projections onto an isolated spectral set. 
 The results of this paper suggest
 then that there is no continuum index (e.g. a Fredholm index) for our model defined via the Fermi projection; if there were such an index, it would be non-constant, continuous and $\ZZ$-valued, a contradiction.
\end{rem}

\begin{rem}
Recently a violation of the bulk-edge correspondence in the continuum was discussed \cite{PhysRevResearch.2.013147,Graf2021}, a further example illustrating that the equivalence between continuum and discrete topological descriptions cannot be assumed in general.
\end{rem}

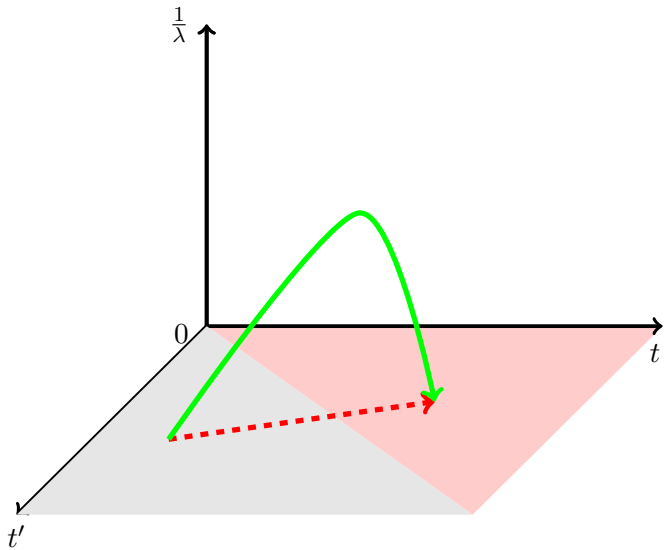
\begin{figure}[h!]
	\centering
\begin{tikzpicture}
	\fill[fill=red!20] (0,0)--(6,0)--(3.5,-2.5);	
	\draw[very thick,line width=0.05cm,->] (0,0) -- (6,0);
	\draw[very thick,line width=0.05cm,->] (0,0) -- (-2.5,-2.5);
	\node[below] at (5.9,-0.1) {$t$};
	\node[below] at (-2.5,-2.5) {$t'$};
	\draw[very thick,line width=0.05cm,->] (0,0) -- (0,4);
	\node[left] at (-0.1,-0.1) {$0$};
	\node[left] at (-0.1,4) {$\frac{1}{\lambda}$};	
	\fill[fill=gray!20] (0,0)--(-2.5,-2.5)--(3.5,-2.5);
	
	\draw[very thick,line width=0.05cm,->] (0,0) -- (0,4);
	\draw[very thick,line width=0.07cm,->,red,dashed] (-0.5,-1.5) -- (3,-1);
	\draw [very thick, line width=0.07cm,green,->] plot [smooth] coordinates {(-0.5,-1.5) (2,1.5) (3,-1)};
	\end{tikzpicture}

	\caption{Schematic of the homotopy we propose. At $\lambda=\infty$, in the $t-t'$ plane, the topological phases are well defined. The grey region corresponding trivial phase ($|t^\prime/t|>1$) and the red region corresponding to the non-trivial phase ($|t^\prime/t|<1$). A deformation along the dashed red path connects trivial to non-trivial phases through a point where the spectral gap closes. The green path, along which $\lambda$ is finite, corresponds to our homotopy of continuum Hamiltonians, along which  the spectral gap does not close.}
	\label{fig:schematic of idea}
\end{figure}

\bigskip
\section{Tight-binding approximation with $\lambda$-dependent lattices}
\label{sec:lambda-dep TB procedure}


Within this section our setting is much more general than the rest of the paper, since here we essentially work in the generality of \cite{Shapiro_Weinstein_2020}: general space dimension $\nu=1,2,3,\dots$ and allowing for a constant magnetic field when $\nu=2$.

Let $\GG\subseteq\RR^\nu$ be a given countable subset obeying the  assumptions of \cite{Shapiro_Weinstein_2020}:
\begin{assumption}
	Strictly positive minimal lattice spacing: $$ a := \inf_{n,m\in\GG:n\neq m}\norm{n-m} > 0\,. $$
\end{assumption}
\begin{assumption}[Separation of coupling length scales] The next-to-nearest-neighbor distances do not converge to $a$ throughout the lattice:
	\begin{align}\label{eq:NNN separation} b := \inf_{n,m\in\GG:\norm{n-m}>a}\norm{n-m} > a\,. \end{align}
\end{assumption}

The chosen lattice $\GG$ does not depend on the asymptotic tight-binding parameter $\lambda$ (to be introduced right below). To model the dependence, we introduce a new $\lambda$-dependent function $$ \ve_\lambda:\GG\to\RR^\nu  $$ which obeys the constraint $$ \lim_{\lambda\to0}\ve_\lambda = 0$$ uniformly in $\GG$, i.e., for some constant $K$, which is independent of $n\in\GG$, \begin{align}\label{eq:decay rate of epsilon} \sup_{n\in\GG}\norm{\ve_\lambda(n)}\le K \lambda^{-\sharp}  \end{align} for some fixed rate $\sharp$ throughout. In the present paper we analyze a model with $\sharp=1$, in principle we expect one needs at least $\sharp\geq1$. The idea is now to place our atoms on the sites given by the lattice $$ \GG_\lambda:=\Set{n+\ve_\lambda(n) | n\in\GG}\,. $$ Having this notation, we still use $\GG\equiv\GG_0$ to index and enumerate the set of lattice points. 

We thus consider the crystal Hamiltonian $$ H_\lambda := (P-b_0\lambda A(X))^2+\lambda^2 \sum_{n\in\GG}v(X-(n+\ve_\lambda(n)))-e\Id\,. $$

Here, $\lambda>0$ models the tight-binding depth (an asymptotic parameter tending to infinity), $b_0 \in \left[0,\infty\right)$ is an $\mathcal{O}(1)$ (in $\lambda$) parameter that controls the relative strength of the magnetic field (or whether it is absent, in which case $b_0=0$), $P,X$ are the momentum and position operators respectively, $A:\RR^\nu\to\RR^\nu$ is the magnetic vector potential for a constant magnetic field $(\nabla\wedge A =  \hat{e}_3)$, $v\in C^\infty(\RR^\nu\to[-1,0])$ is the one-atom electric potential, which we assume to obey the same constraints as in \cite{Shapiro_Weinstein_2020}: the most severe physical constraint is that the spectrum of the one-atom Hamiltonian $$ h^\lambda :=  (P-b_0\lambda A(X))^2+\lambda^2 v(X)$$ remains gapped above its ground state energy as $\lambda\to\infty$. In the magnetic case we also require that $v$ is radial and that the ground state of $h^\lambda$ is, too.

We define $$ \vf_n := \hat{R}^{n+\ve_\lambda(n)}\vf\ $$ where $\hat{R}^x$ is the magnetic translation (see \cite[Eq-n (2.7)]{Shapiro_Weinstein_2020}) and the nearest-neighbor hopping coefficient \begin{align}\label{eq:def of rho} \rho := \lambda^2 \ip{\vf}{v(X) \hat{R}^{a e_1} \vf} > 0 \,. \end{align}
In its definition, $\ve_\lambda$ does not enter.

The tight-binding Hamiltonian $H^{\mathrm{TB}}$ is defined as an operator on $\ell^2(\GG)$ via its matrix-elements \begin{align}\label{eq:TB H} \left(H^{\mathrm{TB}}\right)_{n,m} := \exp\left(\ii b_0 n \wedge m \right)\left[\lim_{\lambda\to\infty} \frac{1}{\rho}\lambda^2\ip{\vf_n}{v(X-n-\ve_\lambda(n))\vf_m}\right]\delta_{a,\norm{n-m}}\qquad(n,m\in\GG)\,. \end{align}
There is a technicality here: if $b_0\neq0$, one needs to take the limit $\lambda\to\infty$ along appropriate sub-sequence so that $\lambda$ does not appear in the oscillating exponential, as in \cite[Def. 2.17]{Shapiro_Weinstein_2020}.

The Gramian is defined, still as an operator on $\ell^2(\GG)$ as, $$ G_{nm} := \ip{\vf_n}{\vf_m} \equiv \ip{\hat{R}^{n+\ve_{\lambda}(n)}\vf}{\hat{R}^{m+\ve_{\lambda}(m)}\vf}\qquad(n,m\in\GG)\,. $$

All the claims of \cite[Section 2.4]{Shapiro_Weinstein_2020} hold also for this Gramian associated with $\GG_\lambda$.

We find (with similar definitions for $\tilde{\vf}$ and $J$ as in \cite[Eq-ns (2.26) and (2.27)]{Shapiro_Weinstein_2020}):
\begin{thm}\label{thm:general lambda-dep tight-binding approx}
	For all $a>a_0$ and $K$ a compact subset of $\sigma(H^{\mathrm{TB}})$, there exists some $\lambda_\star>0$ such that if $\lambda>\lambda_\star,z\in K$, $$ \norm{R_{H/\rho}(z)-J^\ast R_{H^{\mathrm{TB}}}(z)J}\to0\qquad(\lambda\to\infty)\,. $$
\end{thm}
\begin{proof}[Sketch of proof]
	The proof follows the lines of \cite[Theorem 3.1]{Shapiro_Weinstein_2020}. One important aspect is that the minimal lattice spacing of $\GG_\lambda$ now depends on $\lambda$, but is lower bounded as follows:
	$$ \norm{n_\lambda-m_\lambda} \equiv \norm{n-m-\ve_\lambda(n)+\ve_\lambda(m)}\geq \norm{n-m}-2K\lambda^{-\sharp} \geq a-2K\lambda^{-\sharp}\,.  $$ We see that if $\lambda$ is chosen sufficiently large then the minimal lattice spacing of $\GG_\lambda$ may be assumed to be larger than some universal minimal ($\lambda$-independent) constant, say, $\tilde{a}:=\frac{1}{2} a$. Then everywhere where the minimal lattice spacing is used, $\tilde{a}$ should be replaced by $a$ in \cite{Shapiro_Weinstein_2020}. E.g., in the proof of \cite[Prop. 2.15]{Shapiro_Weinstein_2020}, or \cite[Lemma 5.2]{Shapiro_Weinstein_2020}.
	
	Next, one should clarify what role a constraint on $\sharp$ would play. To that end, consider the basic tunneling amplitudes in the model, which are always of the general schematic form $$ \exp\left(-\lambda d(n-\ve_\lambda(n),m-\ve_\lambda(m)\right)\,. $$ Here, $d$ would be the relevant metric of the model: for non-magnetic models this is the Agmon metric which is also the extremal classical Euclidean action \cite{Simon_1984_10.2307/2007072}, for magnetic models this should be determined by the magnetic field, e.g., as $d(x,y)=b_0\norm{x-y}^2$ \cite{FSW_2022,Helffer_Kachmar_22}. 
	
	The analysis in \cite{Shapiro_Weinstein_2020} is built on a hierarchy of exponentially decaying factors, where the largest factor, $\rho$ (see \cref{eq:def of rho}) is of order $\exp(-\lambda d(0,a)$ and all other terms (e.g. correction terms coming from next-to-nearest-neighbor hopping and beyond) are summable and exponentially smaller than $\exp(-\lambda d(0,a)$. If, however, the corrections stemming from $\ve_\lambda$ were much larger, this would spoil the analysis. For example, if $d(0,a)=\norm{a}$ (as is the case in the present paper) and $\ve_\lambda(n)$ would decay to zero like $\lambda^{-1/2}$, then the correction would be of order $\exp(-\sqrt{\lambda}\times\mathrm{length})$ which is much larger than all other scales, including the main term $\rho$. This is the main constraint which sets \cref{eq:decay rate of epsilon}, which, in the present paper implies $\sharp\geq 1$, and in the magnetic case one most likely needs $\sharp\geq2$.
	
	Otherwise, all other theorems go through with no change.
\end{proof}

\section{Our continuum SSH model}
In this section we apply the results of \cref{sec:lambda-dep TB procedure} with the following choices:  define $\nu=1$, $\GG := \ZZ$ (i.e. we work in one dimension) with ($b_0=0$) (no magnetic field). 

We define the $\lambda$-dependent lattice as follows. Let $\alpha\in(-\frac{1}{2},\frac{1}{2})$ and define \eql{ 
	\ve_\lambda(n) := \frac{\alpha}{\lambda}(-1)^{n},\quad n\in\ZZ.
}
The atomic potential wells are centered at the lattice sites: $$ s_n := n +(-1)^n\frac{\alpha}{\lambda},\quad n\in\ZZ, $$
which  are spaced,in-cell and out-of-cell,  at \emph{alternating} distances: \begin{align}\label{eq:din_dout} d_{\mathrm{in}}:=1-\frac{2\alpha}{\lambda},
\quad d_{\mathrm{out}}:=1+\frac{2\alpha}{\lambda}.\end{align}
Now choose positive constants $\wA$ and $\wB$ and 
define \eq{ w_{n} :=  \begin{cases} 
\wA,& n\in 2\ZZ\\
 \wB,& n\in 2\ZZ+1 \,.
 \end{cases}}
 Finally, let $\chi_S$ denote the indicator function of a set $S$.

Our continuum SSH Hamiltonian,  $H^\lambda$, acting in the Hilbert space $\HH := L^2(\RR)$ is given by:
\begin{align} H^\lambda := -\partial_x^2 - \lambda^2 \sum_{j\in\ZZ} \chi_{\left[-\frac{w_j}{2},\frac{w_j}{2}\right]}(X-s_j)\,. \label{eq:our continuum model}\end{align}
 Hence the potential of the Hamiltonian $ H^\lambda $ is a sum of square well potentials centered about all sites in $\Set{s_j}_{j\in\ZZ}$; 
those centered at $even$ sites have width  $\wA$ and those  centered about $odd$ sites have width $\wB$. This class of Hamiltonians has been studied, for example, in \cite{fefferman_topologically_2017}.

The tight-binding regime is obtained when $\lambda$ is taken sufficiently large yet finite. If, after appropriate translation and scaling, $H^\lambda$ converges to a limiting operator on $\ell^2(\ZZ)$ (see \cref{eq:TB H}), the latter is called the tight-binding limit. 

When $\wA=\wB$, each cell contains a symmetric double-well, and we expect the low energy spectrum of this model should be approximated in terms of the two-band SSH model \cref{HSSHk} with in-cell hopping coefficient $\tin$ and out-of-cell hopping coefficient $\tout$, related, respectively, to $\din$ and $\dout$.  This intuition is now made rigorous.

\begin{rem} We have chosen a piecewise constant crystal potential for computational convenience; in this case the hopping coefficient $\rho$, \cref{eq:def of rho}, may be calculated precisely. We believe that our analysis can extended to compactly supported smooth atomic wells.
\end{rem}
Our general scheme will be as follows. Fix $\tin$ and $\tout$ to be such that $0<\tin<\tout$. Let $H^\lambda(\vec{d},\vec{w})$ denote the continuum Hamiltonian with 
\[
\textrm{well-spacing parameters $\vec{d}=(\din,\dout)$ and well-width parameters $\vec{w}=(\wA,\wB)$.}
\]
We also write $\HSSH(\tin,\tout)$ for the tight binding SSH Hamiltonian with hopping parameters $\tin$ and $\tout$ as in \cref{HSSHk}. We shall construct a continuous family of continuum Hamiltonians on $L^2(\RR)$
\begin{equation} [-1,1]\ni\xi\mapsto H^\lambda(\xi)\equiv H^\lambda(\vec{d}_\lambda(\xi),\vec{w}(\xi)) 
\label{homotop}\end{equation}
satisfying the following three properties:
\begin{enumerate}
\item[(H1)] For all $\xi\in[-1,1]$, $H^\lambda(\xi)$ has a spectral gap about zero energy, and furthermore, using that $\tin<\tout$,
   \item[(H2)] For $H^\lambda(-1)$ has tight binding ($\lambda\to\infty$)  limit $\HSSH(\tin,\tout)$ with topological index (Zak phase) equal to $1$ since $|\tin/\tout|<1$ and
    \item[(H3)] For $H^\lambda(+1)$ has tight binding limit ($\lambda\to\infty$) $\HSSH(\tout,\tin)$ (hopping parameters reversed) with topological index equal to $0$, $|\tout/\tin|<1$.
\end{enumerate}

%
%
%
\section{Tight-binding reduction to the discrete model}
Let us set $\wA=\wB=w$ for now. Consider the eigenvalue problem for the \emph{single-well} Hamiltonian,
 $h^\lambda = -\partial_x^2 -\lambda^2 \chi_{[-w/2,w/2]}(x)$. The bound states are oscillating trigonometric functions inside the well and decaying exponentials outside it. Let $\vf^\lambda$ denote the single-well ground state. Denote by $\vf_j$, the ground state of $h^\lambda$, centered at site $s_j$, hence $\vf^\lambda \equiv \vf_0$. The collection of all lattice-translates of $\vf^\lambda$, $\{\vf^\lambda_j\}$, is called the set of {\it atomic orbitals}. For large $\lambda$, it approximately spans the spectral subspace associated with the first two bands of $H^\lambda$ if $\alpha\neq0$ (i.e. $d_{\mathrm{in}}\neq d_{\mathrm{out}}$; see \cref{eq:din_dout}) or the first band if $\alpha=0$.  

A short calculation shows that the ground state energy of this single-site ground state energy $e_0^\lambda$ is given by the implicit equation
\begin{align}
    \sqrt{\lambda^2+e_0^\lambda}\tan\left(\sqrt{\lambda^2+e_0^\lambda}\frac{w}{2}\right) = \sqrt{-e_0^\lambda}
\end{align} whose asymptotic solution is 
\begin{align}\label{eq:single well ground state energy}
    e_0^\lambda \sim -\lambda^2+\frac{\pi^2}{w^2}-\frac{4\pi^2}{w^3}\frac{1}{\lambda}+\mathcal{O}\left(\frac{1}{\lambda^2}\right)\qquad(\lambda\to\infty)\,.
\end{align}  
The ground state wave function is given by 
\begin{equation} \vf^\lambda(x) = A 
\begin{cases}  \cos(\sqrt{\lambda^2+e_0^\lambda}\ x) & x\in[-w/2,w/2] \\
\cos(\sqrt{\lambda^2+e_0^\lambda}\ \frac{w}{2}) \ee^{-\sqrt{-e_0^\lambda}\ (|x|-w/2)} & |x| \geq w/2 
\end{cases} 
\label{vf-as}
\end{equation}
with $$A = \left(\frac{w}{2}\left(1+\mathrm{sinc}\left(w\sqrt{\lambda^2+e_0^\lambda}\right)\right)+\frac{\cos(\frac{w}{2}\sqrt{\lambda^2+e_0^\lambda})}{\sqrt{-e_0^\lambda}}\right)^{-1/2} \sim \sqrt{\frac{2}{w}}-\frac{\sqrt{2}}{w^{3/2}}\frac{1}{\lambda}+\mathcal{O}\left(\frac{1}{\lambda^2}\right)\,.$$ 

Let us define the (still $\lambda$ dependent) Hamiltonian $[ \tilde{H}^\lambda ]$ as an operator on $\ell^2(\ZZ)$ defined by the matrix elements:
\begin{equation}
    [ \tilde{H}^\lambda ]_{j,k} := \langle\vf^\lambda_j,(H^\lambda-e_0^\lambda\Id)\vf^\lambda_k\rangle_{L^2(\RR)}\,. 
    \label{mels}\end{equation}

The dominant matrix elements, giving rise to the hopping terms in the tight binding limit, come from nearest-neighbors
:
\begin{align}
 \rho^\lambda_1 &\equiv \langle\vf_{2j},(H^\lambda-e_0^\lambda\Id)\vf_{2j+1}\rangle \sim -\frac{8\pi^2}{w^3}\frac{1}{\lambda}\exp(-\lambda (\din-3w/2))  \label{rho1}\\
\rho^\lambda_2 &\equiv \langle\vf_{2j-1},(H^\lambda-e_0^\lambda\Id)\vf_{2j}\rangle \sim -\frac{8\pi^2}{w^3}\frac{1}{\lambda}\exp(-\lambda (\dout -3w/2))
.\label{rho2}\end{align}

In fact, even before approximating $\vf^\lambda$ using the ground state energy asymptotic expansion, we can calculate that \begin{align}\label{eq:ratio of hopping coefficients} \rho_2^\lambda/\rho_1^\lambda \sim \exp\left(-\sqrt{-e_0^\lambda}\left(d_{\mathrm{out}}-d_{\mathrm{in}}\right)\right) \end{align} which does not include exponential corrections from NNN wells, but is exact in $\vf^\lambda$ (whose corrections are polynomial in $1/\lambda$).

In the present simple model, these quantities may be estimated by explicit calculation with the 
explicit relations \cref{eq:single well ground state energy} and \cref{vf-as}. In more general settings, e.g. 2D models considered in \cite{Shapiro_Weinstein_2020,FLW17_doi:10.1002/cpa.21735}, lower bounds on these matrix elements are proved; see also \cite{FSW_2022} for the magnetic case.

By \cref{eq:din_dout} we have the "in" and "out" atomic spacings:
\begin{align}\dout = \din + \frac{4}{\lambda}\alpha,\quad \textrm{where $\alpha>0$.}
\label{toflam}\end{align} 
Since $\din<\dout$, we have from \cref{eq:ratio of hopping coefficients}  that $\rho_1^\lambda>\rho_2^\lambda$.

To capture the leading order behavior for $\lambda$ large, we divide \cref{mels} by $\max\left(\Set{\rho_1^\lambda,\rho_2^\lambda}\right)=\rho_1^\lambda$ and let $\lambda$ tend to infinity. This yields an SSH Hamiltonian
 $\HSSH(\tin,\tout)$ acting in $l^2(\ZZ;\CC^2)$
with hopping coefficients:
\begin{align}
    \tin= \rho_1^\lambda/ \rho_1^\lambda = 1\quad\textrm{and}\quad 
     \tout=\lim_{\lambda\to\infty} \rho_2^\lambda/ \rho_1^\lambda =
      \lim_{\lambda\to\infty} e^{-\lambda(\dout-\din)} = e^{-4\alpha}<1=\tin.
\end{align}

Using the same strategy of proof as in \cite{Shapiro_Weinstein_2020} (our setting is only simpler now), the general \cref{thm:general lambda-dep tight-binding approx} reduces to \begin{thm}\label{ssh-thm} There is a partial isometry $J^\lambda:L^2(\RR)\to l^2(\ZZ;\CC^2)$ such that for any $z$ in the resolvent set of $\HSSH(1,e^{-\alpha})$,
	\[ \left[(\rho_1^\lambda)^{-1} (H^\lambda-e_0^\lambda\Id)-z\Id_{L^2}\right]^{-1} \  \textrm{converges to\  $(J^\lambda)^*\left(\HSSH(1,e^{-\alpha})-z\Id_{l^2}\right)^{-1}J^\lambda\ $ as $\lambda\to\infty$}\]
	in the space of bounded linear operators on $L^2(\RR^2)$.
\end{thm}
Since in \cref{ssh-thm} the resulting SSH Hamiltonian has $\tin=1>e^{-4\alpha}=\tout$, by our earlier discussion, this limiting SSH model is topologically trivial (winding number zero). If, rather than
 \cref{toflam}, we take: $\din = \dout + \frac{4}{\lambda}\alpha$, the limiting SSH Hamiltonian is
  $\HSSH(e^{-4\alpha},1)$, which is topologically non-trivial (winding number one).

\section{The homotopy }

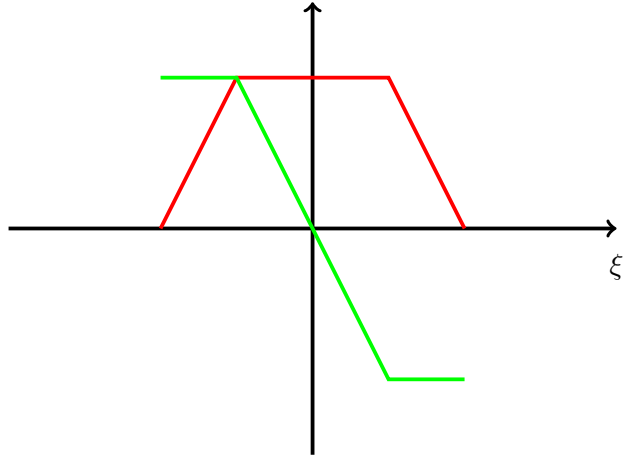
\begin{figure}[h!]
	\centering
	\begin{tikzpicture}[scale=2]
		\draw[very thick,line width=0.05cm,->] (-2,0) -- (2,0);
		\node[below] at (2,-0.1) {$\xi$};
		\draw[very thick,line width=0.05cm,->] (0,-1.5) -- (0,1.5);
		\draw[very thick,line width=0.05cm,-,red] (-1,0) -- (-0.5,1) -- (0.5,1) -- (1,0);
		\draw[very thick,line width=0.05cm,-,green] (-1,1) -- (-0.5,1) -- (0.5,-1) -- (1,-1);
	\end{tikzpicture}

	\caption{Homotopy of the asymmetry parameters $\xi\mapsto\eta(\xi)$ (red) and $\xi\mapsto\delta(\xi)$ (green).}
	\label{fig:homotopy as a function of xi}
\end{figure}
We finally construct the homotopy of continuum Hamiltonians: \[ [-1,1]\ni\xi\mapsto H^\lambda(\xi)=H^\lambda(\vec{d}_\lambda(\xi),\vec{w}(\xi)), \]
 discussed in \cref{homotop}, satisfying properties (H1), (H2) and (H3). In addition to well-separation degrees of freedom
  $\din$ and $\dout$, we make use of the well-width degrees of freedom $\wA$ and $\wB$; we no longer constrain them to be equal, which means that now the tight-binding reduction  \cref{ssh-thm} is no longer applicable as the wells are not identical.
  
We specify the homotopy  by introducing two asymmetry parameters, $\eta$ and $\delta$; the parameter $\eta$ controls the width asymmetry and $\delta$ controls the displacement asymmetry. We take the well-width and well-separation parameters to be of the form:
$$ \vec{w}=(w_{A},w_B) = (w+\eta,w-\eta),\quad  \vec{d}_\lambda= (d_{\mathrm{out}},d_{\mathrm{in}}) = (1+ \delta,1-\delta)\,.  $$ 
We now let $\eta$ and $\delta$ be functions of $\xi$ as follows (see \cref{fig:homotopy as a function of xi}):
$$\eta(\xi) := \beta
\begin{cases}
2\xi+2 & \xi\in[-1,-1/2]\\
1 & \xi\in[-1/2,1/2]\\
-2\xi+2&\xi\in[1/2,1]\end{cases} $$ 
and
$$\delta(\xi) := \frac{2\alpha}{\lambda}
\begin{cases}
1 & \xi\in[-1,-1/2]\\
-2\xi & \xi\in[-1/2,1/2]\\
-1&\xi\in[1/2,1]\end{cases}\,. $$ 
Here $\beta>0$ is a parameter that specifies the maximal asymmetry in the width of the wells. 


%

We claim that, for $\lambda>\lambda_\star$ sufficiently large, $H^\lambda(\vec{d}_\lambda(\xi),\vec{w}(\xi))$ satisfies the following properties:
\begin{enumerate}
\item For all $\xi\in[-1,1]$,  $H^\lambda(\xi)$ has a gap between its first two bands.
\item $H^\lambda(-1)$ has parameters $(\din,\dout)=\left(1+2\frac{\alpha}{\lambda},1-2\frac{\alpha}{\lambda}\right)$, has a spectral gap between its first and second bands, and has the topologically trivial tight binding limit $\HSSH(1,e^{-4\alpha})$
   \item $H^\lambda(+1)$ has parameters $(\din,\dout)=\left(1-2\frac{\alpha}{\lambda},1+2\frac{\alpha}{\lambda}\right)$, has a spectral gap between its first and second bands,  and has the topologically non-trivial tight binding limit $\HSSH(e^{-4\alpha},1)$.
\end{enumerate}

Properties 2) and 3), corresponding to (H2) and (H3),  follow from our discussion of the tight binding limit and remarks in \cref{ssh}. It remains to verify Property 1) for $\lambda>\lambda_\star$.
This is equivalent to the associated Schr\"odinger (Bloch) Hamiltonian, $H^\lambda(k;\xi)-e_0^\lambda\Id$, considered with periodic boundary conditions, having a gap between its two smallest eigenvalues, $\mu^\lambda_1(k;\xi)$ and $\mu^\lambda_2(k;\xi)$, for all $\lambda$ sufficiently large, uniformly in $k\in[0,2\pi]$ and all $\xi\in[-1,1]$.  By the strict monotonicity properties of dispersion curves about $k=0$ and $k=\pi$  \cite{reed1972methods,kuchment2016overview}, it suffices to verify that for all $\lambda>\lambda_\star$ sufficiently large, there is a constant, $c_\lambda>0$, such that:
\begin{align}
   \min_{-1\le\xi\le1}\left( \mu^\lambda_2(0;\xi)-\mu^\lambda_1(0;\xi)\right) \ge c_\lambda\quad {\rm and}\quad 
     \min_{-1\le\xi\le1}\left( \mu^\lambda_2(\pi;\xi)-\mu^\lambda_1(\pi;\xi)\right) \ge c_\lambda.
\label{gap12}\end{align}

\paragraph{Sketch of an analytic argument}
Suppose, to the contrary, that there are  sequences $(\lambda_j)_j\to\infty$ and $(\xi_j)_j\subset[-1,1]$ along which each of the operators $H^{\lambda_j}(\xi_j)-e_0^\lambda\Id$  has a multiplicity two  $k-$pseudo-periodic eigenvalue
 for either $k=0$ (periodic) or $k=\pi$ (anti-periodic). Consider the case $k=0$; the case $k=\pi$ is treated analogously. By compactness of $[-1,1]$, we may pass to a subsequence, $(\lambda_{j_k})\to\infty$ for which $\xi_{j_k}\to\xi_\star\in[-1,1]$. We preclude the cases $\xi_\star=\pm1$ and $|\xi_\star|<1$ separately.

Suppose $\xi_\star=\pm1$. Then, since $w_A(\xi_\star)=w_B(\xi_*)$ and $\alpha>0$, the limiting spectrum (about energy zero) is controlled  by a tight binding Hamiltonian $\HSSH(\tin,\tout)$
 with $\tin\ne \tout$. Our tight binding analysis implies, by passing perhaps to a further subsequence, gives a
strictly positive lower bound the spectral gap of order: 
\begin{align}\label{eq:dw splitting order of magnitude}  \rho_1^\lambda \times \Big|\tin-\tout\Big|= 
\rho_1^\lambda \times \Big|1-e^{-4\alpha}\Big| \ {\rm for}\ \xi_\star=\pm1.  
\end{align} 
This contradicts the assumption of a double eigenvalue (and hence a gap closing) along a sequence.

Now suppose $-1<\xi_\star<1$. The well-widths $w_A(\xi_\star)$ and $w_B(\xi_\star)$, independent of $\lambda$, now differ.  By \cref{eq:single well ground state energy}, the spectral gap opened by this  asymmetry is of order $$ \pi^2\ \Big|\frac{1}{w_A^2(\xi_\star)}-\frac{1}{w_B^2(\xi_\star)}\Big|\ +\ \mathcal{O}\left(\frac{1}{\lambda}\right)\,. $$ 
Hence, this gap opening is of order $1$  quantity for  $\lambda\gg1$. However, the effect of the double-well splitting \cref{eq:dw splitting order of magnitude} is exponentially small in $\lambda$ for $\lambda\gg1$.  Hence, for $\lambda$ sufficiently large, we claim that this effect does not cancel out the order one gap arising from width asymmetry. 
A complete rigorous proof would require justification of the latter assertion.

\section{Numerical analysis and edge modes} 
In this section we complement the above discussion by a numerical study of the homotopy for the infinite continuum bulk  and semi-infinite ``edge'' systems. Our numerical computations of eigenvalues and eigenvectors is based on a  finite-difference method approximation of  $H^\lambda(\xi)$.

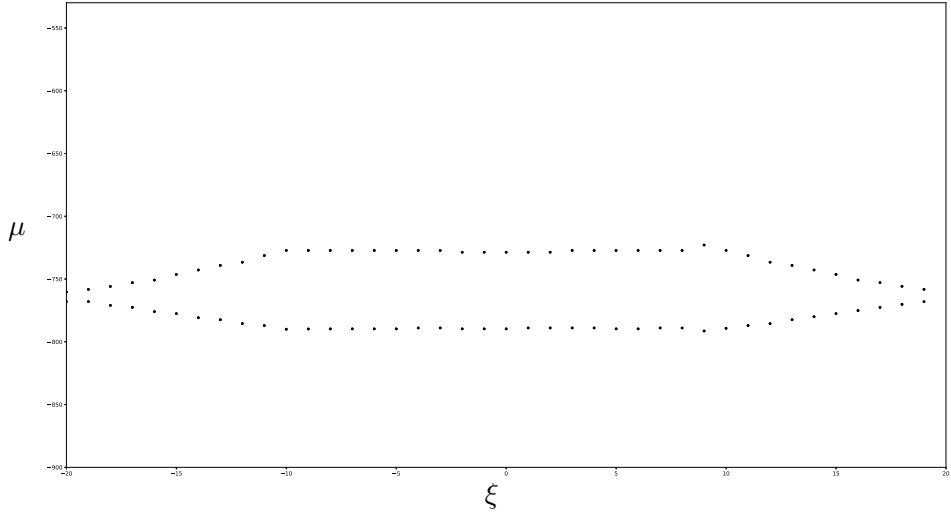
\begin{figure}[h!]
	\centering

\begin{tikzpicture}
    \draw (0, 0) node[inner sep=0] {\includegraphics[scale=0.5,height=8cm]{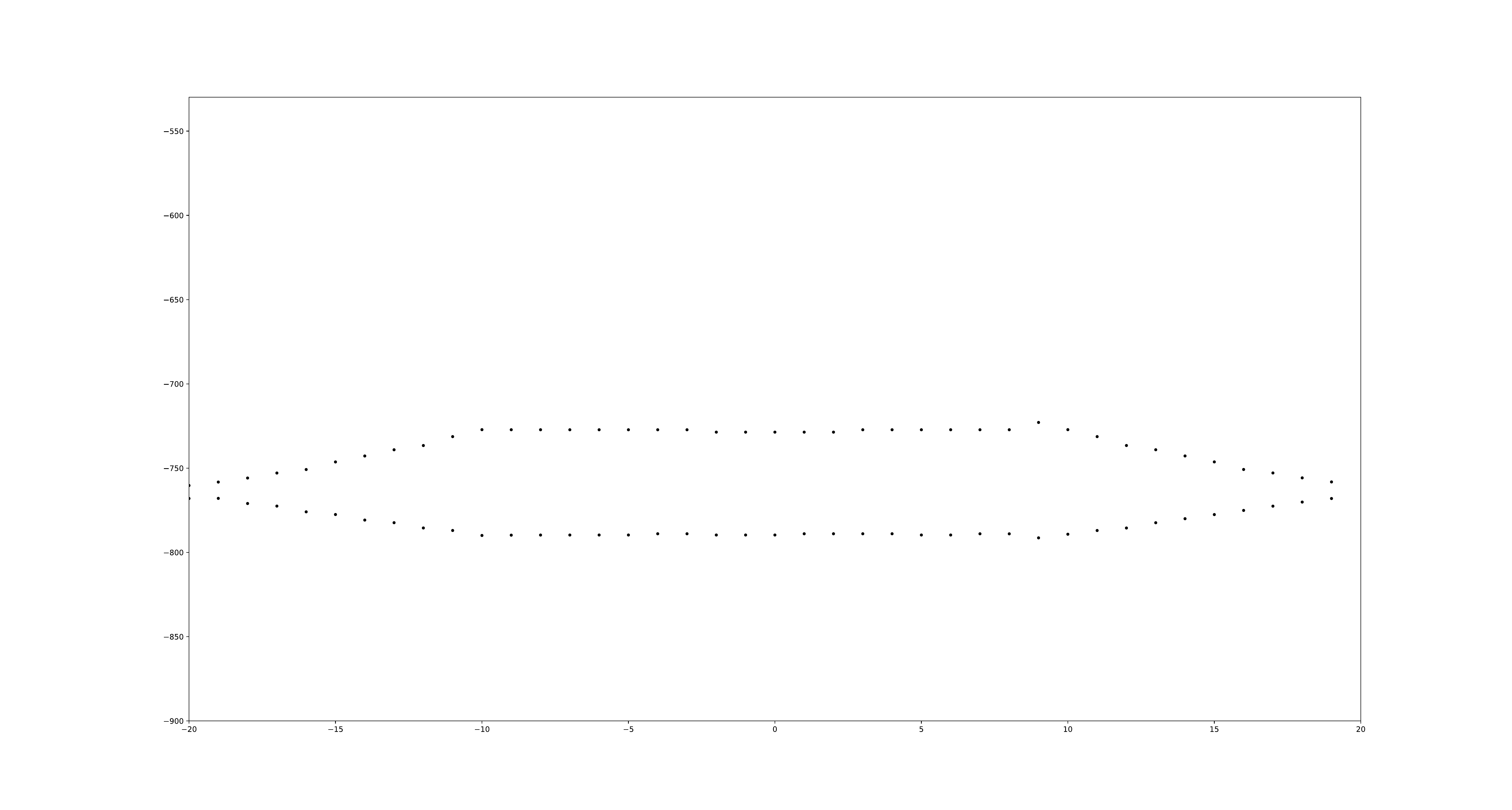}};
    \draw (-6.25, 0) node {$\mu$};
    \draw (0, -3.5) node {$\xi$};
    \draw (0, 3.8) node {$\mu_1(k=0;\xi) < \mu_2(k=0;\xi)$, 2 lowest band dispersion curves of $H^\lambda(k=0,\xi)$};
\end{tikzpicture}
	
	\caption{ Numerical calculation of the two lowest $k=0$  Bloch eigenvalues displaying a gap throughout the homotopy. The results for the two lowest $k=\pi$  Bloch eigenvalues is completely analogous. } 
	\label{fig:energy dispersion vs interpolation parameter}
\end{figure}

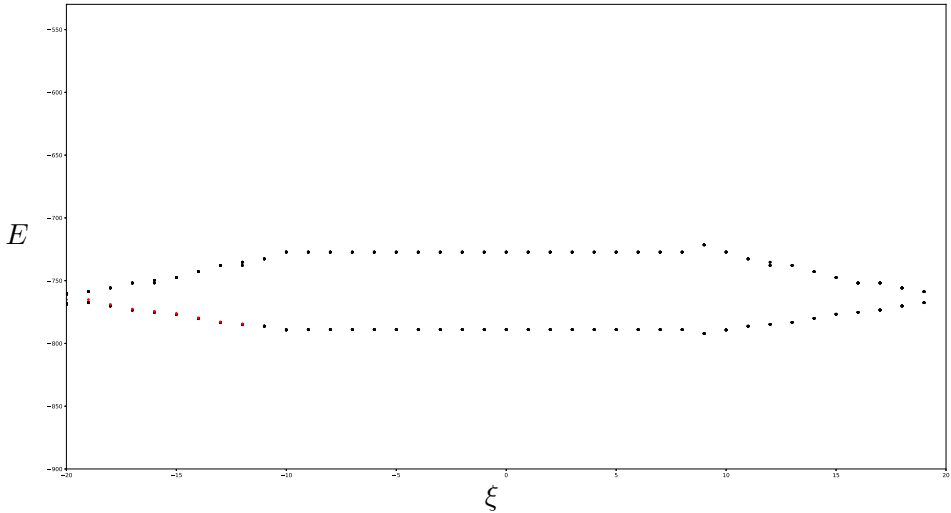
\begin{figure}[h!]
	\centering
    \centering
\begin{tikzpicture}
    \draw (0, 0) node[inner sep=0] {\includegraphics[scale=0.15,height=8cm]{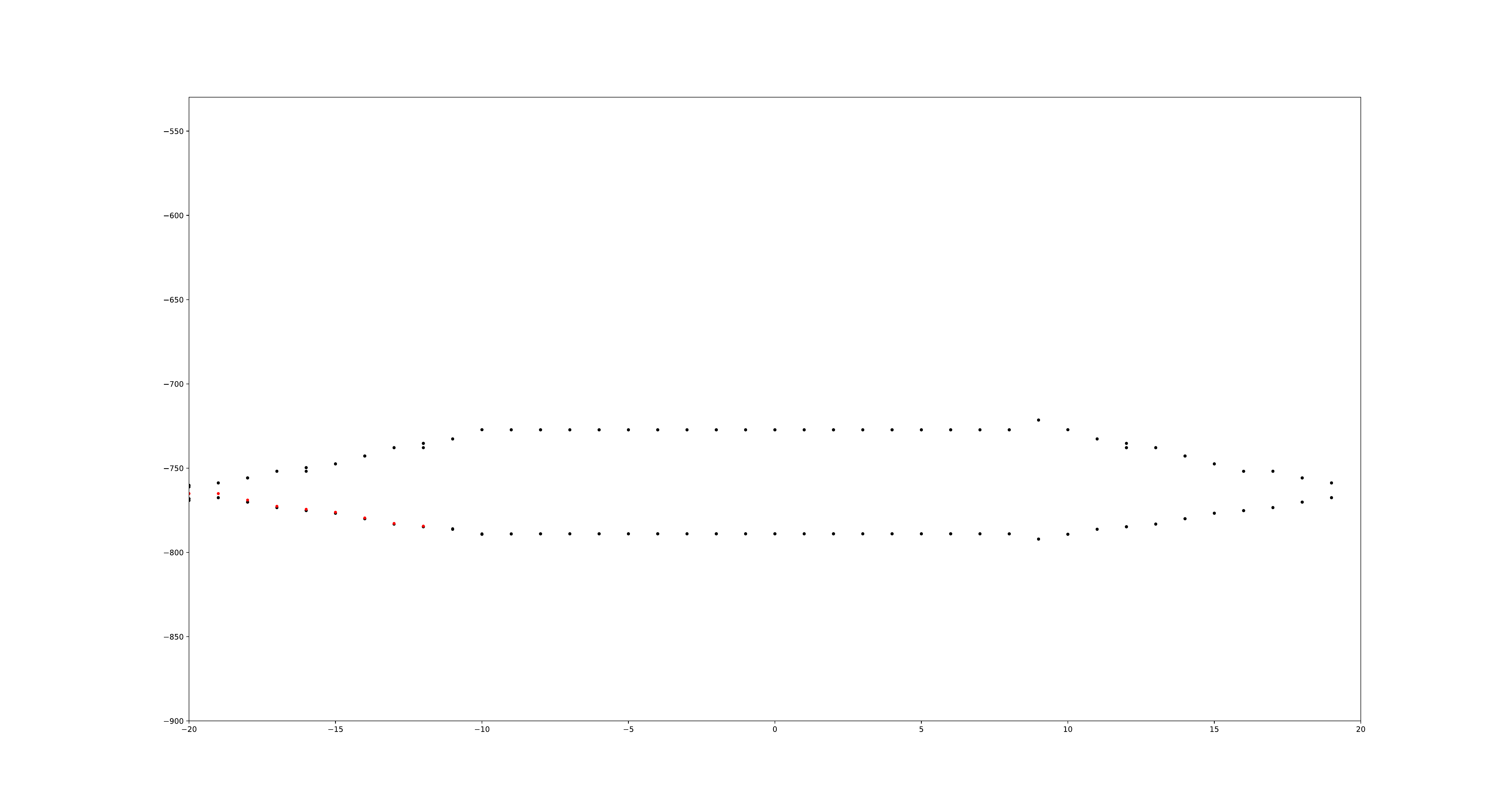}};
    \draw (-6.25, 0) node {$E$};
    \draw (0, -3.5) node {$\xi$};
    \draw (0, 3.8) node {$\mu_1(\xi) < \mu_2(\xi)$, two lowest bulk modes and the edge mode in between (red) of $H^\lambda(\xi)$};
\end{tikzpicture}

	\caption{Numerical calculation of edge modes. Edge mode is enters into the continuous spectrum in the course of the homotopy.}
	\label{fig:edge energy dispersion vs interpolation parameter}
\end{figure}

\begin{figure}[h!]
	\centering
    \centering
\begin{tikzpicture}
    \draw (0, 0) node[inner sep=0] {\includegraphics[angle=-90,origin=c,width=\linewidth]{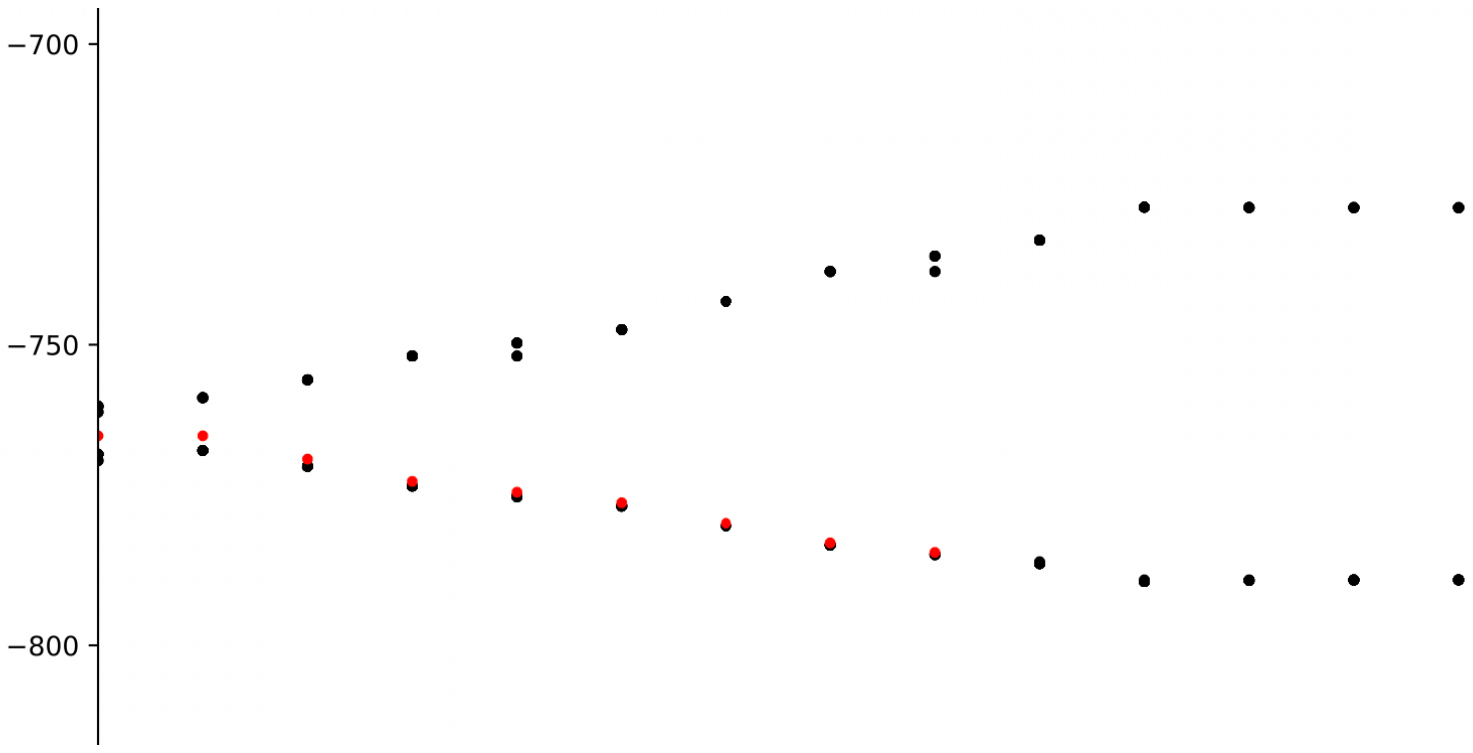}};
\end{tikzpicture}

	\caption{Zoom in on a region of interest in \cref{fig:edge energy dispersion vs interpolation parameter}.}
	\label{fig:zoom in edge dispersion}
\end{figure}

\begin{enumerate}
    \item {\bf Infinite bulk:}
    We exhibit a particular value of $\lambda$ such that for all $\xi\in[-1,1]$, \cref{gap12} holds. 
    Recall that to verify the persistence of the bulk spectral gap throughout the homotopy it is sufficient to prove a strictly positive width gap between the lowest two  $k=0$ (periodic) and $k=\pi$ (anti-periodic) Bloch eigenvalues, which is uniform in $\xi\in[-1,1]$. Consider the case $k=0$. We discretize the bulk Hamiltonian using a mesh of $N=1,000$ equally spaced points over \emph{one} unit cell. The eigenvalues of the resulting matrix eigenvalue problem approximate the bulk spectrum of $H^\lambda(\xi,k=0)$. The parameters $\lambda=30,w=0.2,\beta=0.03,\alpha=3$ were used. In \cref{fig:energy dispersion vs interpolation parameter} we plot the two lowest eigenvalues as a function of the homotopy parameter $\xi$. \cref{fig:energy dispersion vs interpolation parameter} displays a strictly positive gap between these eigenvalues, uniform in $\xi\in[-1,1]$. The $k=\pi$ numerical calculation was performed as well and yielded similar results.

    \item  {\bf Semi-infinite system:} It is well known, via the bulk-edge correspondence \cite{Graf_Shapiro_2018_1D_Chiral_BEC}, that if the Zak phase, or winding number is non-zero, a half-infinite \emph{discrete} system--which is the truncation of the bulk system--has edge modes. Starting with a topological configuration of the discrete system, for which the corresponding continuum problem has an edge mode (for $\lambda$ sufficiently large), we study the motion of this continuum edge mode eigenvalue, for fixed $\lambda$ large,  as the homotopy parameter, $\xi$, varies. As an edge state eigenvalue problem we take
\begin{equation} H^\lambda_\sharp\psi = E\psi,\quad \psi\in L^2(\mathbb{R}),
\label{edge-evp}\end{equation}
where $H^\lambda_\sharp$ denotes the edge Hamiltonian:
\[ H^\lambda_\sharp := -\partial_x^2 - \lambda^2 \sum_{j\ge0} \chi_{\left[-\frac{w_j}{2},\frac{w_j}{2}\right]}(X-s_j),\]
whose potential is supported for $x\ge0$.
We have discretized \eqref{edge-evp}  on the domain $-L\le x \le L$ ($L=10$) and assign 
free boundary conditions, i.e., we use a mesh of $N=20,000$ discrete points, with $20$ unit cells (so each unit cell ahs $1,000$ discrete points, as in the Bloch calculation above). The other parameters are as above: $\lambda=30,\ w=0.2,\ \beta=0.03,\ \alpha=3$. The eigenvectors and eigenvalues of the discretized eigenvalue problem  were calculated. The  inverse participation ratio of the eigenvectors was used to discern between bulk (black) and edge (red) modes. Localized (edge) modes residing in the middle of the system (around $x\approx0$) were plotted; the other localized states were supported at the artificial boundary truncation and are a numerical artifact.  The results of numerical calculations are presented in \cref{fig:edge energy dispersion vs interpolation parameter} and in the magnification \cref{fig:zoom in edge dispersion}. These figures show that the edge mode eigenvalue  gets absorbed in the lower bulk continuous spectrum in the course of the homotopy.
\end{enumerate}


\pagebreak
\bigskip

\noindent\textbf{Acknowledgements:} 
We thank Gian Michele Graf for stimulating discussions and Amir Sagiv for help with numerical simulations. J.S. acknowledges support by the Swiss National Science Foundation (grant number P2EZP2\_184228), as well as support from the Columbia University Mathematics Department and Simons Foundation Award \#376319, while a postdoctoral fellow
during 2018-2019. 
M.I.W. was supported in part by National Science Foundation grants DMS-1412560, DMS-1620418 and DMS-1908657 as well as by the Simons Foundation Math + X Investigator Award \#376319. 
\bigskip

 \bigskip

\begingroup
\let\itshape\upshape
\printbibliography
\endgroup
\end{document}